\documentclass[12pt]{article}
\newlength{\mytopmargin}
\newlength{\myleftmargin}
\usepackage{amsmath,amsthm,amssymb,amsbsy,epsfig,graphicx,color,multicol,subfigure}
\usepackage{array,calc}
\usepackage[enableskew]{youngtab}
\usepackage{rotating}
\usepackage{young,epic}
\usepackage{a4wide,bm}
\usepackage{url}
\usepackage{hyperref}

\newtheorem{lemma}{Lemma}

\usepackage{amsmath,amsfonts,amssymb}

\usepackage{graphicx}

\begin{document}
%

\title{The Golden-Thompson inequality --- historical aspects and random matrix applications}
\author{Peter J. Forrester and Colin J. Thompson }
\date{}
\maketitle
\noindent
\thanks{\small Department of Mathematics and Statistics, 
The University of Melbourne,
Victoria 3010, Australia email:  p.forrester@ms.unimelb.edu.au 
}

\begin{abstract}
The Golden-Thompson inequality, ${\rm Tr} \, (e^{A + B}) \le {\rm Tr} \, (e^A e^B)$ for $A,B$ Hermitian matrices, appeared
in independent works by Golden and Thompson published in 1965. Both of these were motivated by considerations in
statistical mechanics. In recent years the Golden-Thompson inequality has found applications to random matrix theory. In this
survey article we detail some historical aspects relating to Thompson's work, giving in particular an hitherto unpublished
proof due to Dyson, and correspondence with P\'olya. We show too how the $2 \times 2$ case relates to hyperbolic geometry,
and how the original inequality holds true with the trace operation replaced by any
unitarily invariant norm. In relation to the random matrix applications, we review its use in the derivation
of concentration type lemmas for sums
of random matrices due to 
Ahlswede-Winter, and Oliveira, generalizing various classical results.

\end{abstract}

\section{Introduction}
Let $A$ and $B$ be $N \times N$ Hermitian matrices. The Golden-Thompson inequality asserts that
\begin{equation}\label{1}
{\rm Tr} \, (e^{A + B}) \le {\rm Tr} \, (e^A e^B).
\end{equation}
This result has been known since 1964--1965, and has its origins in considerations from statistical mechanics.
Thus Golden, in the paper `Lower bounds for the Helmoltz function' \cite{Go65} proved (\ref{1}) for $A$ and $B$ Hermitian and
non-negative definite. Independently Thompson, in `Inequality with applications to statistical mechanics' 
\cite{Th65}, proved (\ref{1}) without any requirement that the Hermitian matrices $A$ and $B$ be non-negative definite.
In a further remarkable coincidence, also published in 1965 and motivated by statistical mechanics, Symanzik in the
paper `Proof of refinements of an inequality of Feynman' \cite{Sy65} derived (\ref{1}) in the case of
$A = - {d^2 \over d x^2}$, $B = V(x)$ and thus particular Hermitian operators in Hilbert space.

This latter case is easy to illustrate. First, with $H =  - {d^2 \over d x^2} + V(x)$, (\ref{1}) reduces to
\begin{equation}\label{1a}
{\rm Tr} \, e^{-\beta H} \le {1 \over 2 \pi} \int_{-\infty}^\infty d p \, e^{-\beta p^2}  \int_{-\infty}^\infty dx \, e^{-\beta V(x)},
\end{equation}
which in words says that the classical partition function is an upper bound for the corresponding quantum partition function.
If we take $V(x) = x^2$, the integral over position on the RHS is, like the integral over momentum, a Gaussian, while
 the energy levels of the quantum problem are $E_n = 2n + 1$, $n=0,1,\dots$, similarly allowing the LHS to be made explicit to
 give
\begin{equation}\label{1b}
{1 \over \sinh \beta} \le {1 \over \beta}.
\end{equation}

In recent years, renewed prominence has been given to the Golden-Thompson inequality, due to its applicability in
random matrix theory. The latter is the main research area of coauthor Forrester of the present paper. In 1979 Forrester was
a student in the second year linear algebra course of coauthor  Thompson--- the same Thompson as in Golden-Thompson ---
at the University of Melbourne. The inequality (\ref{1}), with the additional information imparted to the class that
Golden was a person rather than an adjective, was presented as enrichment knowledge in one of the lectures.
Upon learning of the renewed interest in (\ref{1}) due to its application in random matrix theory, the
suggestion was then made from Forrester to Thompson of a review article, with the initial intention of familiarising
potential users in random matrix theory with the surrounding mathematics.

Moreover, it is fair to say that the lineage in relation to Thompson's paper \cite{Th65} is rather rich and of
independent interest.
Part of this,
as recorded in the acknowledgements of \cite{Th65}, relates to an until now unpublished  proof of (\ref{1}) by Dyson.
In the early 1960's Dyson wrote a series of seminal papers relating to random matrix theory
and its application to nuclear physics (these were numbered by Roman numerals I up to V, except the
foundational paper `The three fold way', logically number zero in the series and so lacking labelling
by a Roman numeral  \cite{Dy62c}). It is thus a happy coincidence that (\ref{1}) should find
application in this field, and pleasing too that the opportunity to write this article
should have arisen during Freeman Dyson's 90th birthday year.

In Section 2 we detail the mathematics relating to the $2 \times 2$ case of (\ref{1}) (in particular its relation to hyperbolic geometry),
Dyson's proof of (\ref{1}), Thompson's simplification and we comment too on Golden's proof. Thompson's proof
makes use of results of Weyl and P\'olya, and some correspondence with the latter is quoted. The topic of Section 3 is
inequalities of a similar type to (\ref{1}). One such result is that (\ref{1}) holds with the trace operation replaced by any
unitarily invariant norm, and a further relation to hyperbolic geometry of a generalization of this latter result in noted.
The application of (\ref{1}) to deriving operator norm bounds on certain sums of random matrices is given in Section 4.

\section{Proof and generalizations}
\subsection{The 2 by 2 case}
Upon being lead to (\ref{1}) by a different consideration in statistical mechanics than the one mentioned in \cite{Th65} ---
specifically from mathematical structures appearing in the variational Quasi-Chemical Equilibrium Theory of
 superconductivity as observed
during his PhD thesis \cite{Th64} --- Thompson first took up the challenge of proving the $2 \times 2$ case. This case can be
further simplified by the observation that $2 \times 2$ matrices $X$ can be written as the sum of a multiple of the identity,
say $c_X \mathbb I$,
and a traceless matrix $\tilde{X}$.  Since in general $e^{A + B} = e^A  e^B$ when $A$ and $B$ commute, a simple corollary of
the definition of $e^X$ in terms of a power series, we see that ${\rm Tr} \, e^{A + B}  = e^{c_A + c_B}  {\rm Tr} \, e^{{\tilde A} + \tilde B}$
and similarly ${\rm Tr} \, (e^A e^B) = e^{c_A + c_B}  {\rm Tr} \, ( e^{{\tilde A} } e^{\tilde B})$. Thus in the $2 \times 2$ case
it suffices to restrict attention
to $A$ and $B$ traceless matrices in (\ref{1}).

The latter can each be parametrised by three real numbers $a_1, a_2, a_3$ and $b_1, b_2, b_3$ say, by writing
\begin{equation}\label{AB}
A = a_1 \sigma_1 + a_2 \sigma_2 + a_3 \sigma_3, \qquad
B = b_1 \sigma_1 + b_2 \sigma_2 + b_3 \sigma_3,
\end{equation}
where $\sigma_1, \sigma_2, \sigma_3$ are the $2 \times 2$ Pauli matrices
$$
\sigma_1 = \begin{bmatrix} 0 & 1 \\ 1 & 0 \end{bmatrix}, \qquad
\sigma_2 = \begin{bmatrix} 0 & -i \\ i & 0 \end{bmatrix}, \qquad
\sigma_3 = \begin{bmatrix} 1 & 0 \\ 0 & -1 \end{bmatrix}.
$$
An elementary calculation shows $A^2 = || \vec{a} ||^2  \mathbb I$, $|| \vec{a} ||^2 := \sum_{i=1}^3 a_i^2$, and thus
(see e.g.~\cite{Th71})
$$
\exp A = \cosh || \vec{a} || \, \mathbb I + {\sinh ||\vec{a}||  \over || \vec{a} ||} \, A.
$$
From this, and the analogous equation for $B$, after substituting in (\ref{1}) and making essential use
of $A$ and $B$ being traceless to simplify the RHS, we see that our task in establishing (\ref{1})
in the $2 \times 2$ case reduces  to proving that
\begin{equation}\label{1a}
\cosh || \vec{a} + \vec{b} || \le \cosh  || \vec{a} || \, \cosh  || \vec{b} ||  -  \cos \theta
\sinh || \vec{a} || \, \sinh || \vec{b} ||,
 \quad \cos \theta :=  - {  \vec{a} \cdot \vec{b} \over   || \vec{a} || \,  || \vec{b} || }, 
\end{equation}
where as usual $\vec{a} \cdot \vec{b} := \sum_{i=1}^3 a_i b_i$. 

At this stage a connection with hyperbolic geometry is revealed. One recognises the RHS of (\ref{1a}) as $\cosh ||\vec{c}||$, in
a hyperbolic geometry of curvature $-1$, as given by the law of cosines for a triangle of side lengths $|| \vec{a}||,
|| \vec{b}||, ||\vec{c}||$, and angle between $|| \vec{a}||$ and $|| \vec{b}||$ equal to $\theta$ (all measured in
the hyperbolic geometry). So now, after first taking the inverse hyperbolic cosine of both sides and then squaring,
we have reduced (\ref{1a}) to the statement that for a triangle in a hyperbolic  geometry
\begin{equation}\label{1aA}
||\vec{c}||^2 \ge || \vec{a}||^2 + || \vec{b}||^2 - 2|| \vec{a}|| \, || \vec{b}|| \cos \theta.
\end{equation}
The latter is well known \cite{He62}. 

We remark that it is also possible to prove (\ref{1a}) directly. Thus first introduce the parameter $\epsilon$ by the scalings $\vec{a} \mapsto \epsilon \vec{a}$
and  $\vec{b} \mapsto \epsilon \vec{b}$. Then check that the inequality holds term-by-term in like powers of $\epsilon$ (which are all even). The
$\epsilon^0$ and $\epsilon^2$ terms are in fact equalities; from then on one simply needs Cauchy's inequality.

\subsection{Dyson's proof}
At the time of being led to (\ref{1}), Thompson  had just began a postdoc at UCSD and Dyson was visiting there on sabbatical. The result (\ref{1}) was shown
to Dyson in late 1964 as a conjecture for $N \ge 3$, and soon after he produced a proof.
Working from the original notes, we record this proof in the following working.

\begin{lemma}\label{L1}
For any square matrices $X$, $Y$
$$
| {\rm Tr} \, (XY) |^2 \le {\rm Tr} \, (X^\dagger X) \,  {\rm Tr} \, (Y^\dagger Y) .
$$
\end{lemma}

\noindent
Proof. \quad This is Cauchy's inequality. \hfill $\square$

\medskip
\begin{lemma}\label{L2}
Let $P$ be any product of $2n$ factors which may be $X$ or $X^\dagger$ in any order.
Then
$$
| {\rm Tr} \, P | \le {\rm Tr} \, (X X^\dagger)^n.
$$
\end{lemma}

\noindent
Proof. \quad Among all the products $P$ choose one for which $| {\rm Tr} \, P|$ is the greatest. If $P = (X X^\dagger)^n$ or
$(X^\dagger X)^n$ there is nothing to prove. Otherwise there appears somewhere in $P$ a pair of consecutive factors $X$
or $X^\dagger$. Permute the factors cyclically so that the two $X$ or $X^\dagger$ appear in positions $n, n+1$.
Now write $P = QR$, where $Q$ is the product of the first $n$ factors, $R$ the remainder. 

By Lemma \ref{L1},
$$
| {\rm Tr} \, P|^2 \le {\rm Tr} \, (Q^\dagger Q) \, {\rm Tr} \, (R^\dagger R) .
$$
But $Q^\dagger Q := P'$, $R^\dagger R := P''$ are both products of the same type as $P$, which means that
$$
| {\rm Tr} \, P'| \le | {\rm Tr} \, P|, \qquad | {\rm Tr} \, P''| \le | {\rm Tr} \, P|.
$$
Therefore
$$
| {\rm Tr} \, P | = | {\rm Tr} \, P' | = | {\rm Tr} \, P''|.
$$
Now the number of neighbour-pairs $X X^\dagger$ in $P,P',P''$ may be denoted by $k,k',k''$ respectively. We count the last and
the first factors as neighbours in the trace. We have then
$$
k' + k'' = 2k+1, \: 2k+2
$$
according as the first and the last factors in $P$ were different or the same. In any case {\it at least one} of $P'$ and $P''$ has more
neighbour-pairs $X X^\dagger$ than $P$.

Now apply this argument to the product $P$ with maximum $| {\rm Tr} \, P|$ and with the maximum number of neighour-pairs $X X^\dagger$.
We have a contradiction unless $P = (X X^\dagger)^n$ or $(X^\dagger X)^n$. This proves the lemma.
 \hfill $\square$
 
 \medskip
 \begin{lemma}\label{L3}
 For any two Hermitian matrices $A$ and $B$
 $$
 {\rm Tr} \, (A^{2^k} B^{2^k}) \ge {\rm Tr} \, (AB)^{2^k}.
 $$
 \end{lemma}
 
 \noindent
Proof. \quad Taking $X = AB$, $X^\dagger = BA$ in Lemma \ref{L2}, we have
\begin{equation}\label{u1}
| {\rm Tr} \, (AB)^{2n} | \le {\rm Tr} \, (ABBA)^n = {\rm Tr} \, (A^2 B^2)^n.
\end{equation}
Next take $X = A^2 B^2$, so that
$$
| {\rm Tr} \, (AB)^{4n} | \le {\rm Tr} \, (A^2B^2B^2A^2)^n = {\rm Tr} \, (A^4 B^4)^n.
$$
Repeating this argument leads at once to Lemma \ref{L3}. 
 \hfill $\square$
 
 \medskip
 \noindent
 {\bf Proof of (\ref{1}).} \quad Write in Lemma \ref{L3} $A' = (1 + 2^{-k}A)$ for $A$,   $B' = (1 + 2^{-k}B)$ for $B$ and take the limit
 $k \to \infty$.   \hfill $\square$
 
 \medskip
 Before we proceed to give Thompson's \cite{Th65} simplification of the above working, for the purpose of historical links,
 suppose that in Lemma \ref{L3} we write $A' = e^{A/2^k}$ for $A$ and $B' = e^{B/2^k}$ for $B$. Then taking the limit
 $k \to \infty$ on the RHS asks us to compute $\lim_{k \to \infty} ( e^{A/2^k} e^{B/2^k} )^{2^k}$. For this one can
 use the Lie-Trotter formula \cite{LE88,Tr59},
 \begin{equation}\label{LT}
 \lim_{n \to \infty} (e^{A/n} e^{B/n} )^n = e^{A + B},
 \end{equation}
 which according to some sources \cite{CHMM78} was found by Lie in 1875.
 
 \subsection{Thompson's proof}
 We have just seen that the last step in Dyson's proof makes use of a fundamental matrix identity. Thompson's \cite{Th65} simplification similarly
 calls upon fundamental matrix relations, now between the eigenvalues $|\lambda_1| \ge \cdots  \ge |\lambda_N|$ of a matrix $X$ and
 its singular values $\mu_1 \ge \cdots \ge \mu_N$ (i.e.~positive square roots of the eigenvalues of $X^\dagger X$), due to Weyl \cite{We49}.
 
  \begin{lemma}\label{L4}
  Let $w(x)$ be an increasing function of positive argument $x$, with the further property that $w(e^\xi)$ is a convex function of $\xi$. One has
  \begin{equation}\label{2.6}
  \sum_{i=1}^k w(\mu_i) \ge \sum_{i=1}^k w(|\lambda_i|), \qquad k=1,\dots,N.
  \end{equation}
 \end{lemma}
 
 Taking $w(x) = x^{2s}$, $(s=1,2,\dots)$, $k = N$, (\ref{2.6}) reads
 \begin{equation}\label{H}
 \sum_{i=1}^N \mu_i^{2s} \ge \sum_{i=1}^N |\lambda_i |^{2s} \ge | \sum_{i=1}^N \lambda_i^{2s} |,
 \end{equation}
 where the final relation is just the triangle inequality, or equivalently
 \begin{equation}\label{W2}
 {\rm Tr} \, (X^\dagger X)^s \ge | {\rm Tr} \, X^{2s} |, \qquad s=1,2,\dots.
 \end{equation}
The inequality (\ref{W2}) is given as Lemma \ref{L1} in  \cite{Th65}. Armed with (\ref{W2}), one now has an alternative to Dyson's
Lemma \ref{L2} to deduce the result of Lemma \ref{L3} and thus (\ref{1}). The argument is precisely the same
\cite{Th65}.

 \subsection{Golden's proof}
 As in the above workings, Golden (a physical chemist at Brandeis  during the years 1951--1981 \cite{Sa95})
in \cite{Go65}  first establishes the inequality of Lemma \ref{L3}, then concludes the argument through use of
 (\ref{LT}). And as in Dyson's strategy, explicit use too is made of a Cauchy inequality, though not that in Lemma \ref{L1}, but rather
 $$
 0 \le {\rm Tr} \, X^2 \le {\rm Tr} \, X X^\dagger
 $$
 (cf.~(\ref{W2}) with $s=1$) which is restricted to the case that $X$ is non-negative definite.
 
 In fact in the case of non-negative definite matrices $A,B$ the inequality (\ref{u1}) underlying the proof
 of Lemma \ref{L3} can be generalized to the Araki-Lieb-Thirring
inequality  \cite{Ar90,LT76}
$$
 0 \le {\rm Tr} \, (A^{1/2} B A^{1/2} )^{rs}  \le {\rm Tr} \, (A^{r/2} B^r A^{r/2} )^s, \qquad r \ge 1, \: s> 0
 $$
 (take $r=2$, $s = n$ \cite{BPL04}).
 
 \subsection{The George P\'olya link}
 Weyl's result Lemma \ref{L4} is deduced in two steps. The first is to establish the inequalities 
 $\prod_{i=1}^k |\lambda_i| \le  \prod_{i=1}^k \mu_i$. The second is to apply the following lemma,
 with $a_i = \log \mu_i$, $b_i = \log |\lambda_i|$ and $w(x) = \omega(e^\xi)$.
 
  \begin{lemma}\label{L5}
  Given two sequences of real numbers $a_1,\dots, a_m$ and $b_1,\dots,b_m$ such that $b_1 \ge b_2 \ge \cdots \ge b_m$ and
  \begin{equation}\label{i1}
  b_1 + \cdots + b_q \le a_1 + \cdots + a_q, \qquad q=1,\dots,m
  \end{equation}
  the inequality
  $$
  \omega(b_1) + \cdots + \omega(b_m) \le \omega(a_1) + \cdots + \omega(a_m)
  $$
  holds for any convex increasing function $\omega(x)$.
  \end{lemma}
  
 P\'olya \cite{Po50} pointed out that this lemma can be deduced from an inequality of Hardy, Littlewood and P\'olya \cite{HLP28},
  and rediscovered a few years later by Karamata \cite{Ka32}. The latter differs from Lemma \ref{L5} in that the last inequality in
  (\ref{i1}), namely the case $q=m$, is required to be an equality, and that $\omega(x)$ is assumed to be convex but need not be
  increasing.
  
  After isolating (\ref{W2}) as a simplification to Dyson's Lemma \ref{L2} in the pathway to the proof of (\ref{1}), and thus the
  role played by earlier results of Weyl and P\'olya, Thompson wrote to George P\'olya (who was
  actually a Great Uncle by marriage) with this proof, and an enquiry about the possibility of it having already appeared in the
  literature. We quote from the subsequent reply:
  
  \medskip
  {\tt 
  \begin{flushright} Zurich, Feb.~5 '65 \end{flushright}
  
  \medskip
  \noindent
  Dear Colin,

  \qquad \qquad I was much pleased to receive your letter. The theorem
  $$
  {\rm tr} \, (e^A e^B) \ge {\rm tr} (e^{A+B)})
  $$
  is new to me, and I find it very neat. It, is true, I did not look up the \\  literature --- I had,
  or rather I am still having, a bad spell of sciatica \\ which makes very difficult for me to sit in the
  library \& I have very few \\ books in my hotel room. I must even confess that I remember only
  vaguely \\ the notes by Weyl and by myself you are quoting. $\dots$ We should arrive in
 \\ Stanford March 13. In Stanford, I can consult Loewner who knows much more \\ about matrices
  than myself. With my sciatica, the trip may present serious complications --- but we shall see.
  $\dots$
  
 \quad \quad  Your Uncle George
 
 \noindent
 Please give my best regards to Prof.~Dyson.}
 
 \medskip
 A letter dated the Rockefeller Institute, Tuesday, March 2 (1965) from P\'olya to Thompson,
contains  some follow up on previous knowledge of (\ref{1}): ``I asked several mathematicians about that
 nice result (\ref{1}) e.g.~Richard Brauer (Harvard) Radamacher (Philadelphia) etc.~but nobody knew it."
 And then, again sent from the Rockefeller Institute, on March 5 (1965): ``I was just shown
 Physical Review vol.~137 No.~4B 22 Feb.~1965 p.~B1127." The latter is of course Golden's article.
 
 Dyson suggested to Thompson that with the simplified proof and additional results that he should
 submit anyhow (by himself), with an appropriate cover letter to the Editor, and indeed the paper was
 accepted.

  \section{Related inequalities and generalizations}
 Let $\phi(X)$ be a continuous real valued function of the eigenvalues of $X$, and suppose
 \begin{equation}\label{4}
 \phi((X^\dagger X)^s) \ge | \phi(X^{2s}) |, \qquad s=1,2,\dots
 \end{equation}
 We know from Lemma \ref{H} that
 \begin{equation}\label{4.2}
 \phi(X) := \sum_{i=1}^k | \lambda_i |
 \end{equation}
 is an explicit example. As noted by Thompson \cite{Th71}, the proof of (\ref{1}) via Lemma \ref{L3} now yields
 \begin{equation}\label{4.1}
 \phi( e^{A + B}) \le \phi(e^A e^B).
 \end{equation}
 An equivalent viewpoint on (\ref{4.1}) in the case that $\phi$ is given by
 (\ref{4.2}) is that
 \begin{equation}
 e^{A + B} \prec_w e^A e^B,
 \end{equation}
 where for $X,Y$ Hermitian $X \prec_w Y$ (in words $Y$ weakly majorizes $X$) means that
 $\sum_{i=1}^k \lambda_i(X) \le \sum_{i=1}^k \lambda_i(Y)$ ($k=1,\dots,N$). 
 Choosing $\phi(X) = \sum_{i=1}^N | \lambda_i |^p$, $p \ge 1$, one has in the case that $X$ is Hermitian and
 non-negative
 definite $\phi(X) =  ({\rm Tr} \, (X X^\dagger )^{p/2} )= (|| X ||_p)^p$ (here $|| \cdot ||_p$ denotes the
 so-called Schatten $p$-norm) and we see that (\ref{4.1}) gives
  \begin{equation}\label{5}
 || e^{A + B} ||_p  \le  || e^A e^B ||_p.
 \end{equation}
 We remark that (\ref{5}) can equivalently be written \cite{HP93}
   \begin{equation}\label{5a}
   {\rm Tr} \,  e^{A + B}  \le {\rm Tr} \, (e^{pB/2} e^{pA} e^{pB/2} )^{1/p}.
   \end{equation}
 In the case $p \to \infty$ (\ref{5})  was first established by Segal \cite{Se69}, while Leonard \cite{Le71} and
 Thompson \cite{Th71} showed that it  holds for general unitary invariant norms.
 
 A generalisation of (\ref{5}) is \cite{Ba02x}
 \begin{equation}\label{Sn}
 || A - B|| \le || \log ( e^{-B/2} e^A e^{-B/2} ) ||
\end{equation}
valid for $A,B$ Hermitian and the norm $|| \cdot ||$ unitarily invariant. In the case  $|| \cdot || = || \cdot ||_2$ and
with $A$ and $B$ positive definite (\ref{Sn}) reads
\begin{equation}\label{Sn1}
 || A - B||_2 \le \Big ( \sum_{i=1}^N \Big ( \log \lambda_i (e^A e^{-B}) \Big )^2 \Big )^{1/2} =: \delta_2 (e^A, e^B),
 \end{equation}
 where $\{\lambda_i(X)\}$ denotes the eigenvalues of $X$.
 The significance of $\delta_2$ is that it is the distance associated with the invariant Riemannian metric
 $d s^2 = \Big ( {\rm Tr} \, (A^{-1} d A) \Big )^2$. 
 Since $||A||_2 = \Big ( \sum_{i=1}^N ( \log \lambda_i (e^A) )^2 \Big )^{1/2}$, it follows from the definition in (\ref{Sn1}) that
 $||A||_2 = \delta_2(e^A, \mathbb I)$. This in turn allows (\ref{Sn1}), after squaring both sides, to be interpreted as
 a law of cosines for triangles in a hyperbolic geometry \cite{Mo55, Ko73,Ba02x,BH06}, as already seen in (\ref{1aA}).
 
 On another front, using essentially the same working as used by Thompson to obtain (\ref{4.1}), Cohen et al.~\cite{CFKK82}
 generalized (\ref{4.1}) to
 \begin{equation}\label{4.1a}
 | \phi(e^{A + B}) | \le \phi \Big ( e^{(A + A^\dagger)/2}   e^{(B + B^\dagger)/2}  \Big ),
 \end{equation}
where it is no longer required that $A,B$ are Hermitian. Note that the special case $B = 0$, $\phi$ given by (\ref{4.2}) with
$k=1$ yields \cite[Corollary 5]{CFKK82}
 \begin{equation}\label{4.1b}
 \lambda_1\Big ( {1 \over 2} (A + A^\dagger) \Big ) \ge {\rm Re} \, \lambda_1 (A).
 \end{equation}
 We remark that in distinction to (\ref{5}) with $p=\infty$ (\ref{4.1a}) for $\phi$ given by (\ref{4.2}) with
$k=1$ is not a statement about matrix norms. Thus one has that $\lambda_1(A)| = ||A||_\infty$ for
$A$ Hermitian, but more generally $|\lambda_1(A) | \le ||A||_\infty$. This latter property has been
used in \cite{De78} to prove (\ref{5}) with $p=\infty$.
 
 A basic question related to (\ref{1}) is the condition for equality. It has already been remarked that when
 $A,B$ commute, $e^{A} e^{B} = e^{A + B}$, and so the former is a sufficient condition for equality.
 It can be proved that this is also a necessary condition \cite{So92}. One way to see this is to expand
 (\ref{1}) to increasing higher orders of the matrix products \cite[p.~253]{Ta12}. To third order
 both sides agree, while the fourth order terms differ by a term proportional to Tr$\, [A,B]$. Another
 basic question relates to extending (\ref{1}) or the underlying inequality from Lemma \ref{L3}, to involve
 three or more matrices.
 In the original paper \cite{Th65} a counter-example involving $3 \times 3$ real symmetric matrices
  to the obvious generalization of the latter, $| {\rm Tr} \, (ABC)^{2^k} | \le {\rm Tr} \, (A^{2^k} B^{2^k} C^{2^k})$,
  is given. And a counter-example to the obvious generalization of (\ref{1}) to three matrices
    \begin{equation}\label{4.1d} 
   {\rm Tr} \, (e^{A+B+C}) \le | {\rm Tr} \, (e^A e^B e^C) | ,
   \end{equation} 
   involving the $2 \times 2$ Pauli 
  matrices, is given in \cite{Th71}. In relation to this latter setting, a result of Lieb \cite{Li73} gives
  \begin{equation}\label{4.1c} 
  {\rm Tr} \, (e^{A+B+C})\le \int_0^\infty {\rm Tr} \, \Big (e^A (t + e^{-C})^{-1} e^B (t+ e^{-C})^{-1} \Big ) \, dt,
 \end{equation} 
 which one sees reduces to (\ref{1}) when $C=0$. Moreover, (\ref{4.1c}) shows that if $[B,C]=0$, then
 (\ref{4.1d}) (without the need for the absolute value on the RHS) holds true. However, since in this circumstance
 $e^{B+C} = e^{B} e^{C}$ as noted in the first paragraph of Section 2.1, then (\ref{4.1d}) is just (\ref{1}) with $B$
 replaced by $B+C$.

\section{Applications to random matrix theory}

The first application of the Golden-Thompson inequality to random matrix theory was made by Ahlswede and Winter
\cite{AW02}. This contribution was later popularized by it featuring  in a blog article of Tao
\cite{Ta10}, and this blog article in turn became a section in the book on random matrices of the latter
\cite{Ta12}.

To motivate this line of research, let us recall \cite[Preface]{Fo10} that random matrix theory has its historical origin in the
work of Hurwitz \cite{Hu98}, who computed the volume form of a general unitary matrix parametrized in terms of Euler angles.
This work was influential  in the development of the concept of the Haar measure on classical groups, and more generally
locally compact topological groups, in the work of Haar, von Neumann, Weyl, Weil, Seigel and others
\cite[pg.~143]{MR58}. In the case of positive definite matrices the Haar measure is induced by the invariant Riemann metric
noted below (\ref{Sn1}). One consequence of the work of Hurwitz is that the precise probability density function on the
four non-zero, non-unit entries, of a set of $N(N-1)/2$ unimodular matrices can be read off, such that when multiplied
in an appropriate order, and further multiplied by the scalar $e^{i \alpha_0}$, $-\pi \le \alpha_0 < \pi$ chosen uniformly at
random, the corresponding unitary matrix corresponds to sampling $U(N)$ with Haar measure \cite{ZK94}.

There are alternative ways to sample $U(N)$ with Haar measure \cite{Me07}. One is to begin with an $N \times N$ complex standard Gaussian
matrix (all entries independent with standard complex Gaussian entries) $X$, and to form the matrix $V := (X^\dagger X)^{-1/2} X$.
Another is to apply  the Gram-Schmidt algorithm to the columns of $X$. Granted these facts, a natural question is to quantify
properties of $U \in U(N)$ in common with $X$. A celebrated result of Dyson \cite{Dy62a} is that the local eigenvalue statistics
of $V$ and ${1 \over 2}(X + X^\dagger)$ (the latter in the so-called bulk; see e.g.~\cite[Ch.~5]{Fo10}) agree in the $N \to \infty$ limit. At the 
level of the distribution of the entries, the top $k \times k$ block of $U$, $U_k$ say, has distribution with probability
density proportional to \cite{ZS99}
$$
\Big ( \det ( \mathbb I_k - U_k^\dagger U_k) \Big )^{N-2k}.
$$
Thus for $N \to \infty$, and with $k$ fixed, the probability density function of $G_k := \sqrt{N} U_k$ is proportional to
$e^{- G_k^\dagger G_k}$, and hence the entries of $G_k$ are independent complex standard  Gaussians in distribution.
The analogue of this result for Haar distributed real orthogonal matrices, telling us that in the limit $N \to \infty$
the top $k \times k$ block, when scaled by
$\sqrt{N}$, is distributed as a real Gaussian matrix of independent standard Gaussian entries, is known to hold
for $k = {\rm o}(\sqrt{N})$ in the variation norm \cite{Ji06}.

Related to this interplay between Gaussian and unitary random matrices is the question of quantifying the orthogonality
between columns of a standard complex Gaussian matrix. This has application, for example, to the problem of bringing a convex body
to near-isotropic position \cite{Ru99}, and that of analyzing low rank approximations of matrices \cite{RV07,HMT11}.
Denote by $X_{N,k}$ the first $k$ columns of  an $N \times N$ complex standard Gaussian matrix  $X$, and form the
empirical covariance matrix
\begin{equation}\label{S}
\Sigma := {1 \over N}  X_{N,k}^\dagger  X_{N,k}.
\end{equation}
Since all entries are independent, with zero mean and unit standard derivation, we see immediately that $\mathbb E \, \Sigma = \mathbb I_k$.
To measure the distance from the mean, introduce the operator norm $|| X||_{\rm op} =  \mathop{{\rm sup}}_{|| \vec{x} || = 1} || X \vec{x} ||$.
This is related to the Schatten $p$-norm by $|| X ||_{\rm op} =  || X ||_{\infty}$ and so for $X$ Hermitian we have
\begin{equation}\label{SP}
|| X ||_{\rm op}  = {\rm max} \, \{ - \lambda_{\rm min}(X), \lambda_{\rm max}(X) \}.
\end{equation}

For fixed $k$ and large $N$ it follows from the central limit theorem applied to the elements of $\Sigma$ that
$|| \Sigma - I_k ||_{\rm op}$ goes to zero at a rate proportional to  $1/\sqrt{N}$. Of interest is a large deviation bound relating to
$|| \Sigma - I_k ||_{\rm op}$, which is valid for general $N$ and $k$.
We first observe
\begin{equation}\label{S3}
\Sigma = {1 \over N} \Big [ \sum_{p=1}^N \bar{x}_{i,p} x_{p,j} \Big ]_{i,j=1,\dots,k} =
 {1 \over N}  \Big [ \sum_{p=1}^N \bar{x}_{i,p} x_{p,j} \Big ]_{i,j=1,\dots,k} 
 =  {1 \over N}   \sum_{p=1}^N  (\vec{X}^{(p)})^\dagger \vec{X}^{(p)} ,
 \end{equation}
 where $ \vec{X}^{(p)}$ is the row vector formed by the $p$-th row of $X_{N,k}$.  This exhibits $\Sigma$ as the average of 
 independent and identically distributed (rank one) random matrices. 
Thus the task at hand can be interpreted as a noncommutative analogue of classical questions relating to the distribution
of the scalar average $S :={1 \over N} \sum_{p=1}^N s_p$ for $\{s_p\}$ identical and independently distributed. 
Specifically, if we take as our task that of bounding ${\rm Pr}( || \Sigma -   \mathbb I_k ||_{\rm op} > \epsilon)$ then we are seeking
a non-commutative analogue of the Chernoff inequality  (large deviation formula) from classical probability theory
\begin{equation}\label{C}
{\rm Pr} \, (s_1 + \cdots + s_N \ge \epsilon) \le {\rm max} \, (e^{-\epsilon^2/4}, e^{-\epsilon \sigma / 2}),
\end{equation}
where $\epsilon > 0$, and the $s_i$ are independent, identically distributed  random variables taking
values in $[-1,1]$ with mean zero and variance $\sigma^2/N$.

For this task, we begin by making use of (\ref{SP}) to observe
\begin{equation}\label{rf}
{\rm Pr}( || \Sigma -  \mathbb I_k ||_{\rm op} > \epsilon) \le  {\rm Pr}( \lambda_{\rm max}(\Sigma -  \mathbb I_k ) > \epsilon) +
{\rm Pr}(- \lambda_{\rm min}(\Sigma -  \mathbb I_k ) > \epsilon) .
\end{equation}
We now consider the first term on the RHS of (\ref{rf}). Bernstein's trick of effectively exponentiating the
statement $ || \Sigma -  \mathbb I_k ||_{\rm op} > \epsilon$ without changing its probability tells us that for any $c>0$
$$
 {\rm Pr}( \lambda_{\rm max}(\Sigma -  \mathbb I_k ) > \epsilon)  =
 {\rm Pr}( e^{c \lambda_{\rm max}(\Sigma -  \mathbb I_k )} > e^{c \epsilon}) =
 {\rm Pr}(\lambda_{\rm max}(e^{c(\Sigma -  \mathbb I_k )}) > e^{c \epsilon}).
 $$
 But
 \begin{equation}\label{J}
 \lambda_{\rm max}(e^{c(\Sigma -  \mathbb I_k )}) \le {\rm Tr}  \, e^{c (\Sigma -  \mathbb I_k )}
 \end{equation}
  and so
 \begin{equation}\label{rf1}
  {\rm Pr}( \lambda_{\rm max}(\Sigma -  \mathbb I_k ) > \epsilon)  \le  {\rm Pr}( {\rm Tr} \, e^{c \lambda_{\rm max}(\Sigma -  \mathbb I_k )}  > e^{c \epsilon}) \le
  e^{-c \epsilon} {\mathbb E} \, ({\rm Tr} \, e^{c (\Sigma - \mathbb I_k)}),
 \end{equation} 
  where the final bound follows from Chebyshev's inequality. The analogous argument applied to the second term on the
  RHS of (\ref{rf}) gives
   \begin{equation}\label{rf2}
  {\rm Pr}( -\lambda_{\rm min}(\Sigma -  \mathbb I_k ) > \epsilon)  \le
  e^{-c \epsilon} {\mathbb E} \, ({\rm Tr} \, e^{-c (\Sigma - \mathbb I_k)}).
 \end{equation} 
  
  So far the working has mimicked that of one of the standard proofs of (\ref{C}) (see e.g.~\cite{Ta12}).
  To proceed further requires a new idea, and this was forthcoming in the work of Ahlswede and Winter \cite{AW02},
  by way of application of the Golden-Thompson inequality (\ref{1}) to bound the average on the RHS of (\ref{rf1}).

\begin{lemma}\label{LAW}
For any $\mu \in \mathbb R$,
\begin{equation}\label{GTE}
\mathbb E ( {\rm Tr} \, e^{\mu (\Sigma -  \mathbb I_k)}) \le k \prod_{p=1}^N || \mathbb E (e^{\mu (( \vec{X}^{(p)})^\dagger \vec{X}^{(p)} -  \mathbb I_k)/N})
||_{\rm op}.
\end{equation}
\end{lemma}

 \noindent
Proof. \quad Write $\Sigma -  \mathbb I_k =: \Sigma^{(N)}$ and $\Sigma^{(N)} = \sum_{p=1}^N S^{(p)}$ so that
$S^{(p)} := (( \vec{X}^{(p)})^\dagger \vec{X}^{(p)} -  \mathbb I_k)/N$. Then, since $ \Sigma^{(N)} =   \Sigma^{(N-1)}  + S^{(N)}$ we have
$$
\mathbb E ( {\rm Tr} \,   e^{\mu \Sigma^{(N)}}  ) =  \mathbb E ( {\rm Tr} \,   e^{\mu \Sigma^{(N-1)} + \mu S^{(N)}}  ) .
$$
Applying the Golden-Thompson inequality (\ref{1}) to the RHS gives
\begin{equation}\label{W1}
\mathbb E ( {\rm Tr} \,   e^{\mu \Sigma^{(N)}}  ) \le   \mathbb E ( {\rm Tr} \,   e^{\mu \Sigma^{(N-1)} } e^{ \mu S^{(N)}}  ) .
\end{equation}

Each of the $S^{(p)}$ are independent, so taking the expectation with respect to $S^{(N)}$, and furthermore making use of the
fact that $\mathbb E$ are Tr commute (to be able to take advantage of this fact explains the
use of (\ref{J})), we see that (\ref{W1}) can be rewritten to read
\begin{align}\label{4.29}
\mathbb E ( {\rm Tr} \,   e^{\mu \Sigma^{(N)}}  ) & \le    \mathbb E_{S^{(1)},\dots,S^{(N-1)}}
{\rm Tr} (  e^{\mu \Sigma^{(N-1)} } \mathbb E_{S^{(N)}} \,  e^{ \mu S^{(N)}}  )\nonumber \\
& \le ||   E_{S^{(N)}} \,  e^{ \mu S^{(N)}}  ||_{\rm op}
 E_{S^{(1)},\dots,S^{(N-1)}}
{\rm Tr} (  e^{\mu \Sigma^{(N-1)} } ),
\end{align}
where the final inequality follows from the general fact that for $A$ positive definite ${\rm Tr} \, AB \le ||B||_{\rm op}
{\rm Tr} A$. Applying (\ref{4.29}) recursively, with base case $N=1$ for which ${\rm Tr} \, e^{\mu \Sigma^{(0)}} = {\rm Tr} \, \mathbb I_k = k$,
gives (\ref{GTE}). \hfill $\square$

\medskip
Substituting (\ref{GTE}) in (\ref{rf1}) and (\ref{rf2}) and recalling (\ref{rf}) gives
$$
{\rm Pr} \, ( || \Sigma - \mathbb I_k ||_{\rm op} > \epsilon) \le k \, e^{- \epsilon c}
\Big (   \prod_{l=1}^N || \mathbb E (e^{ c S^{(l)}}) ||_{\rm op} +
 \prod_{l=1}^N || \mathbb E (e^{ - c  S^{(l)}}) ||_{\rm op} \Big ).
 $$
 The remaining task is to bound $||\mathbb E (e^{ c  S^{(l)}}) ||_{\rm op} $, and then to minimize the RHS with respect to
 $c>0$. This is straightforward; the details can be found in e.g.~\cite{Ve08}. 
 The final result, under the simplifying assumption that $|| S^{(l)} || \le 1$ for each $l=1,\dots,N$ is \cite{Ru99}
 \begin{equation}\label{RU}
 {\rm Pr}( || \Sigma -  \mathbb I_k ||_{\rm op} > \epsilon) \le k \,{\rm max} \, ( e^{-\epsilon^2/(4 \sigma^2)}, e^{- \epsilon / 2}),
 \end{equation}
 where $\sigma^2 := \sum_{l=1}^N ||{\rm Var} \,  S^{(l)}||_{\rm op}$. This is a noncommutative matrix analogue
 of the Chernoff inequality (\ref{C}).

 There is a significant refinement of the Ahlswede-Winter argument due to Oliveira \cite{Ol10a,Ol10b}. One setting where it shows itself
 is in bounding the operator norm of the random matrix sum $Z^{(N)} := \sum_{p=1}^N \epsilon_p A^{(p)}$ where the
 $A^{(p)}$ are fixed $d \times d$ Hermitian matrices while $\{\epsilon_p \}$ are independent Rademacher (uniformly distributed on $[-1,1]$) or
 standard Gaussian random variables. In accordance with the argument leading to (\ref{rf1}), the key task is to bound
 $ \mathbb E ( {\rm Tr} \,   e^{\mu Z^{(N)}}  )  $.
 
 \begin{lemma}
 In terms of the above notation, for all   $\mu \in \mathbb R$,
 \begin{equation}\label{OB}
 \mathbb E ( {\rm Tr} \,   e^{\mu Z^{(N)}} )  \le {\rm Tr} \, \Big ( e^{\mu^2 \sum_{p=1}^N \epsilon_p (A^{(p)})^2} \Big ).
\end{equation} 
 \end{lemma}

 \noindent
Proof. \quad For $j=1,\dots,N$ define
 \begin{equation}\label{DD}
 D^{(j)} = D^{(0)} + \sum_{p=1}^j \Big ( \mu \epsilon_p A^{(p)} - {1 \over 2} \mu^2 A^{(p)} \Big ),
\end{equation} 
with $D^{(0)} :=   {1 \over 2} \mu^2  \sum_{p=1}^N  (  A^{(p)}  )^2$. 

From these definitions
$$
 \mathbb E ( {\rm Tr} \,   e^{\mu D^{(N)}}  )  =   \mathbb E \, {\rm Tr} \,
 e^{D^{(N-1)} + \mu \epsilon_N A^{(N)} - \mu^2 (A^{(N)})^2/2}.
 $$
 Making use of the Golden-Thompson inequality (\ref{1}) on the RHS, then proceeding as in the derivation of (\ref{4.29})
 shows
 \begin{align}\label{DD3}
  \mathbb E ( {\rm Tr} \,   e^{\mu D^{(N)}}  )  & \le || \mathbb E _{ \epsilon_N} ( e^{\mu \epsilon_N A^{(N)} - \mu^2 (A^{(N)})^2/2} ) ||_{\rm op}
  {\rm Tr} \, (e^{\mu D^{(N-1)}}) \nonumber \\
  & = 
  || e^{-\mu^2  (A^{(N)})^2/2}  \mathbb E_{ \epsilon_N} ( e^{\mu \epsilon_N A^{(N)} }) ||_{\rm op}
  {\rm Tr} \, (e^{\mu D^{(N-1)}}) ,
  \end{align}
  where in obtaining the equality, use has been made of $-\mu^2 (A^{(N)})^2/2$ and $\mu \epsilon_N A_N$ commuting.
  
  For $\epsilon_N$ a Rademacher or standard Gaussian random variable, explicit computation reveals
  \begin{equation}\label{DD1}
  || e^{- \mu^2 (A^{(N)})^2/2} \mathbb E_{\epsilon_N} (e^{\mu \epsilon_N A^{(N)}}) ||_{\rm op} =
  || f(A^{(N)}) ||_{\rm op}
  \end{equation}
  where $0 \le f(t) \le 1$ for all $t \in \mathbb R$ and thus (\ref{DD1}) is bounded by 1. This fact allows us to deduce from (\ref{DD3}) the recurrence
  relation
   \begin{equation}\label{DDN}
  \mathbb E ( {\rm Tr} \,   e^{\mu D^{(N)}}  )   \le       \mathbb E ( {\rm Tr} \,   e^{\mu D^{(N-1)}}  ) .
 \end{equation}
Applying (\ref{DDN}) recursively a total of $N$ times and noting $D^{(N)} = Z^{(N)}$ gives (\ref{OB}).
\hfill $\square$

\medskip
Note that a direct adaptation of the proof of Lemma \ref{LAW} would give
$$
\mathbb E ( e^{\mu Z^{(N)}}) \le d e^{\mu^2 \sum_{\mu=1}^N || (A^{(p)})^2||_{\rm op}}.
$$
This is demonstrated in \cite{Ol10b} to always be weaker than (\ref{OB})

The Chernoff inequality (\ref{C}) is one of a large number of probabilistic bounds relating to sums of scalar random
variables. We have seen that the Golden-Thompson inequality (\ref{1}) allows for the derivation of a matrix
generalization (\ref{RU}). It turns out that (\ref{OB}) is key to deriving a bound for $(\mathbb E \, || Z^{(N)} ||^p)^{1/p}$,
which is the matrix generalization of the classical Khintchine inequality \cite{Ol10b}. Recently Tropp \cite{Tr10,Tr12}
has shown that a result of Lieb \cite{Li73} relating to convex trace functions offers a powerful alternative
to the Golden-Thompson inequality in the derivation of matrix generalization of further probabilistic bounds.
On the other hand, as commented in \cite{Tr12}, the approach of Oliveira \cite{Ol10a} using the Golden-Thompson
inequality to establish a matrix generalization of Freedman's inequality (a martingale version of Bernstein's inequality)
has the advantage of of extending
to the fully noncommutative setting \cite{JZ12}.

We conclude by quantifying the values of the sides of (\ref{1}) when averaged over suitable matrix ensembles. We restrict attention
to the case $N = 2$ with $A$ and $B$ traceless as in (\ref{AB}), and the real numbers $a_1,\dots,b_3$ all standard Gaussians. We read off
from the two sides of  (\ref{1a}) that
\begin{align*}
\mathbb E \, {\rm Tr} \, (e^{A+B}) = \Big ( {1 \over 2\pi} \Big )^{3} 
\int_{\mathbb R^6} e^{- || \vec{a}||^2/2 - || \vec{b}||^2/2 } \cosh ||\vec{a} + \vec{b}|| \, d \vec{a} \, d \vec{b} \\
\mathbb E \, {\rm Tr} \, (e^{A}e^B) = \Big ( {1 \over 2\pi} \Big )^{3} 
\int_{\mathbb R^6} e^{- || \vec{a}||^2/2 - || \vec{b}||^2/2 } \cosh ||\vec{a}||  \cosh || \vec{b}|| \, d \vec{a} \, d \vec{b} 
\end{align*}
(the term involving $\cos \theta$ on the RHS of (\ref{1a}) averages to zero). Making use of polar coordinates to
evaluate the integrals shows
\begin{equation}\label{R}
{ \mathbb E \, {\rm Tr} \, (e^{A}e^B) \over \mathbb E \, {\rm Tr} \, (e^{A+B}) } = {4 \over 3}.
\end{equation}

An obvious question is the behaviour of the ratio (\ref{R}) as a function of the matrix size $N$ for $A,B$ from the Gaussian
unitary ensemble, for example. The analogous question for  the inequality (\ref{4.1b}) can readily be answered in the case
that $A$ is a standard real or complex Gaussian matrix. Then well known results for the largest eigenvalues in the Gaussian
orthogonal and unitary ensemble, and for the spectral radius of the real and complex Ginibre ensemble, (see e.g.~\cite{Fo10}) tell us that
$$
\lim_{N \to \infty} { \mathbb E \, \lambda_1 \Big ( {1 \over 2} (A + A^\dagger) \Big ) \over 
 \mathbb E \, {\rm Re} \, \lambda_1 (A) }  = \sqrt{2}.
 $$
 
 \section*{Acknowledgements}
 The work of PJF was supported by the Australian Research Council. We would also like to  thank Freeman Dyson
 for encouraging correspondence and for allowing us to reproduce the proof of his ingenious Lemma 2.


\begin{thebibliography}{10}

\bibitem{AW02}
R.~Ahlswede and A.~Winter, \emph{Strong converse for identification via quantum
  channels}, IEEE Trans. Inform. Theory \textbf{48} (2002), 569--579.

\bibitem{Ar90}
H.~Araki, \emph{On an inequality of {L}ieb and {T}hirring}, Lett. Math. Phys.
  \textbf{19} (1990), 167--170.

\bibitem{BPL04}
N.~Bebiano, J.~da~Providencia, and R.~Lemos, \emph{Matrix inequalities in
  statistical mechanics}, Linear Algebra Appl. \textbf{376} (2004), 265--273.

\bibitem{Ba02x}
R.~Bhatia, \emph{On the exponential metric increasing property}, Linear Algebra
  Appl. \textbf{375} (2003), 211--220.
  
\bibitem{BH06}
R.~Bhatia and J.~Holbrook, \emph{Noncommutative geometric means}, Math. Intelligencier
 \textbf{28} (2006), 32--39.


\bibitem{CHMM78}
A.J. Chorin, T.J. Hughes, M.F. McCracken, and J.E. Marsden, \emph{Product
  formulas and numerical algorithms}, Comm. Pure Appl. Math \textbf{31} (1978),
  205--256.

\bibitem{CFKK82}
J.E. Cohen, S.~Friedland, T.~Kato, and F.P. Kelly, \emph{Eigenvalues
  inequalities for products of matrix exponentials}, Linear Algebra Appl.
  \textbf{45} (1982), 55--95.
  
  \bibitem{De78}
  P.A. Deift, \emph{Application of a commutation formula},
  Duke Math. J.  \textbf{45} (1978), 267--310.

\bibitem{Dy62a}
F.J. Dyson, \emph{Statistical theory of energy levels of complex systems
  {III}}, J. Math. Phys. \textbf{3} (1962), 166--175.

\bibitem{Dy62c}
\bysame, \emph{The three fold way. {Algebraic} structure of symmetry groups and
  ensembles in quantum mechanics}, J. Math. Phys. \textbf{3} (1962),
  1199--1215.

\bibitem{Fo10}
P.J. Forrester, \emph{Log-gases and random matrices}, Princeton University
  Press, Princeton, NJ, 2010.

\bibitem{Go65}
S.~Golden, \emph{Lower bounds for {H}elmholtz function}, Phys. Rev.
  \textbf{137} (1965), B1127--B1128.

\bibitem{HMT11}
N.~Halko, P.G. Martinsson, and J.A. Tropp, \emph{Finding strucuture with
  randomness: probabilistic algorithms for constructing approximate matrix
  decompositions}, SIAM Review \textbf{53} (2011), 217--288.

\bibitem{HLP28}
G.H. Hardy, J.E. Littlewood, and G.~P\'olya, \emph{Some simple inequalities
  satisfied by convex function}, Messenger Math. \textbf{58} (1928/29),
  145--152.

\bibitem{He62}
S.~Helgason, \emph{Differential geometry and symmetric spaces}, Academic, New
  York, 1962.

\bibitem{HP93}
F.~Hiai and D.~Petz, \emph{The {G}olden-{T}hompson trace inequality is
  complemented}, Linear Alg. Appl. \textbf{1993} (1993), 153--185.

\bibitem{Hu98}
A.~Hurwitz, \emph{{\"Uber} die {C}omposition der quadratischen {F}ormen von
  beliebig vielen {V}ariabeln}, Nachr. Ges. Wiss. G\"ottingen (1898), 309--316.

\bibitem{Ji06}
T.~Jiang, \emph{How many entries of a typical orthogonal matrix can be
  approximated by independent normals?}, Ann. Probab. \textbf{34} (2006),
  1497--1529.

\bibitem{JZ12}
M.~Junge and Q.~Zheng, \emph{Noncommutative martingale deviation and Poincar\'e
  type inequalities with applications}, arXiv:1211.3209, 2012.

\bibitem{Ka32}
J.~Karamata, \emph{Sur une in\'egalit\'e relative aux fonctions convexes},
  Publ.~Math.~Univ.~Belgrade \textbf{1} (1932), 145--148.

\bibitem{Ko73}
B.~Kostant, \emph{On convexity, the {W}eyl group and {I}wasawa decomposition},
  Ann. Sci Ecole Norm. Sup \textbf{6} (1973), 413--460.

\bibitem{Le71}
A.~Lenard, \emph{Generalization of the {G}olden-{T}hompson inequality ${\rm tr}
  \, e^{A + B} \le {\rm tr} (e^a e^b)$}, Indiana Univ. Math. J. \textbf{21}
  (1971), 457--467.

\bibitem{LE88}
S.~Lie and F.~Engel, \emph{Theorie der transformationsgruppen}, 3 vols.
  Teubner, Leipzig, 1888.

\bibitem{Li73}
E.H. Lieb, \emph{Convex trace functions and the {W}igner-{Y}anase-{D}yson
  conjecture}, Adv. Math \textbf{11} (1973), 267--288.

\bibitem{LT76}
E.H. Lieb and W.~Thirring, \emph{Inequalities for the moments of the
  eigenvalues of the {S}chr\"odinger {H}amiltonian and their relation to
  {S}obolev inequalities}, Studies in Mathematical Physics: essays in honor of
  {V}alentine {B}argmann (B~Simon E.H.~Lieb and A.~Wightman, eds.), Princeton
  University Press, Princeton, NJ, 1976.

\bibitem{MR58}
A.M. Macbeath and C.A. Rogers, \emph{Siegel's mean value theorem in the
  geometry of numbers}, Math.~Proc. Camb. Phil. Soc. \textbf{54} (1958),
  139--151.
  
\bibitem{Me07}
 F. Mezzadri, \emph{How to generate random matrices from the classical groups
}, Notices AMS \textbf{54} (2007),  592--604.
 

\bibitem{Mo55}
G.D. Mostow, \emph{Some new decomposition theorems for semisimple groups},
  Memoirs Amer. Math. Soc. \textbf{14} (1955), 31--54.

\bibitem{Ol10a}
R.I. Oliveira, \emph{Concentration of the adjacency matrix and of the
  {L}aplacian in random graphs with independent edges}, arXiv:0911.0600.

\bibitem{Ol10b}
\bysame, \emph{Sums of random {H}ermitian matrices and an inequality by
  {R}udelson}, Elec. Commun. Prob. \textbf{15} (2010), 203--212.

\bibitem{Po50}
G.~P\'olya, \emph{Remarks on Weyl's note ``Inequalities between the two kinds
  of eigenvalues of a linear transformation"}, Proc. Natl. Acad. Sci.
  \textbf{36} (1950), 49--51.

\bibitem{Ru99}
M.~Rudelson, \emph{Random vectors in the isotropic position}, J. Funct. Anal.
  \textbf{164} (1999), 60--72.

\bibitem{RV07}
M.~Rudelson and R.~Vershynin, \emph{Sampling from large matrices: an approach
  through geometric functional analysis}, Journal of the ACM \textbf{54}
  (2007), Article 21.



\bibitem{Sa95}
A.B. Sachar, \emph{Brandeis University: A host at last}, University Press of
  New England, Hanover, NH, 1995.

\bibitem{Se69}
I.~Segal, \emph{Notes toward the construction of nonlinear relativistic quantum
  fields {III}}, Bull. Amer. Math. Soc. \textbf{75} (1969), 1390--1395.

\bibitem{So92}
W.~So, \emph{Equality cases in matrix exponential inequalties}, SIAM J. Math.
  Anal. \textbf{1992} (1992), 1154--1158.

\bibitem{Sy65}
K.~Symanzik, \emph{Proof of refinements of an inequality of {F}eynmann},
  J.~Math.~Phys. \textbf{6} (1965), 1155--1156.

\bibitem{Ta10}
T.~Tao, \emph{The {G}olden-{T}hompson inequality},
  terrytao.wordpress.com/2010/07/15/the-golden-thompson-inequality, 2010.

\bibitem{Ta12}
\bysame, \emph{Topics in random matrix theory}, Graduate Studies in
  Mathematics, Vol.~132, American Mathematical Society, Providence, RI, 2012.

\bibitem{Th64}
C.J. Thompson, \emph{Investigations in the theory of superconductivity}, Ph.D.
  thesis, University of New South Wales, 1964.

\bibitem{Th65}
\bysame, \emph{Inequality with applications in statistical mechanics},
  J.~Math.~Phys. \textbf{6} (1965), 1812--813.

\bibitem{Th71}
\bysame, \emph{Inequalities and partial orders on matrix spaces}, Indiana Univ.
  Math. J \textbf{21} (1971), 469--480.

\bibitem{Tr10}
J.A. Tropp, \emph{User-friendly tail bounds for sums of random matrices},
  Found. Comput. Math. \textbf{12} (2012), 389--434.

\bibitem{Tr12}
\bysame, \emph{User-friendly tools for random matrices: an introduction},
  http://users.cms.caltech.edu/\~{}jtropp/notes/Tro12-User-Friendly-Tools-NIPS.pd%
f, 2012.

\bibitem{Tr59}
H.~Trotter, \emph{On the product of semi-groups of operators}, Proc. Amer.
  Math. Soc. \textbf{10} (1959), 545--551.
  
  \bibitem{Ve08}
R.~Vershynin, \emph{A note on sums of independent random matrices after
  {A}hlswede-{W}inter},
  www-personal.umich.edu/\~{}romanv/teaching-group/ahlswede-winter.pdf, 2009.

\bibitem{We49}
H.~Weyl, \emph{Inequalities between the two kinds of eigenvalues of a linear
  transformation}, Proc. Natl. Acad. Sci. \textbf{35} (1949), 408--411.

\bibitem{ZK94}
K.~Zyczkowski and M.~Kus, \emph{Unitary random matrices}, J. Phys. A
  \textbf{27} (1994), 4235--4245.
  
\bibitem{ZS99}
K.~Zyczkowski and H.~J Sommers, \emph{Truncations of random unitary matrices}, J. Phys. A
  \textbf{33} (2000), 2045--2057.


\end{thebibliography}

\providecommand{\bysame}{\leavevmode\hbox to3em{\hrulefill}\thinspace}
\providecommand{\MR}{\relax\ifhmode\unskip\space\fi MR }
\providecommand{\MRhref}[2]{%
  \href{http://www.ams.org/mathscinet-getitem?mr=#1}{#2}
}
\providecommand{\href}[2]{#2}

\end{document}